\documentclass[12pt,letterpaper]{article}
\usepackage{amsmath}
\usepackage{amssymb}
\usepackage{amsfonts}
\usepackage{bbm}
\usepackage{graphicx,epsf}
\usepackage{epstopdf}
\usepackage{enumerate}
\usepackage{caption}
\usepackage{natbib}
\usepackage{bm}
\usepackage{color}
\usepackage{hyperref} 

\newcommand{\gs}{\geqslant}
\newcommand{\ls}{\leqslant}
\newcommand{\by}{\boldsymbol y}
\newcommand{\bC}{\boldsymbol C}
\newcommand{\bT}{\boldsymbol T}
\newcommand{\bZ}{\boldsymbol Z}

\begin{document}

\title{Discussion of ``The power of monitoring''}

\author{Jakob Raymaekers,
				Peter J. Rousseeuw, and
				Iwein Vranckx\\KU Leuven,
				Belgium}
  
\date{February 4, 2018}

\maketitle
\thispagestyle{empty}

{\bf Abstract.} 
This is an invited comment on the discussion
paper ``The power of monitoring: how to make 
the most of a contaminated multivariate sample''
by A. Cerioli, M. Riani, A. Atkinson and
A. Corbellini that will appear in
{\it Statistical Methods \& Applications}.

\vspace{1cm}

We would like to congratulate 
Cerioli, Riani, Atkinson and Corbellini 
(henceforth CRAC) 
on their well-written and lavishly illustrated 
exposition about the usage and benefits of 
monitoring, and thank the editors for inviting 
us to comment on this interesting work.

\section{The problem of nearby 
         contamination}
The leading example in the paper
is the geyser (Old Faithful) dataset.
From the scatterplot of this bivariate 
dataset we see that it consists of two
clusters, the smaller of which contains
about 30\% to 35\% of the observations.
If one interprets the smaller cluster as 
contamination this is a relatively high 
contamination level, though  
it should not be prohibitive since the 
estimators considered in the paper can 
be tuned to a breakdown value well 
above 35\%.
However, the contamination happens to
lie quite close to the inlying data. 
Having a large fraction of contamination
located fairly close by makes the geyser
data particularly challenging, as 
illustrated by CRAC. 
If we use a scatter estimator with 
a breakdown
value of e.g. 40\%, replacing any 35\%
of clean data by data points positioned
anywhere cannot completely destroy
the scatter matrix (in the sense of 
making its first eigenvalue arbitrarily
large or its last eigenvalue arbitrarily
close to zero), but that does not imply
that the scatter matrix will have a
small bias. Indeed, it is known that
the bias of the estimators under study 
is the largest for nearby contamination,
as shown by \cite{Hubert:Shapebias}.

\begin{figure}[!ht]
\centering
\includegraphics[width = 0.7\linewidth]
      {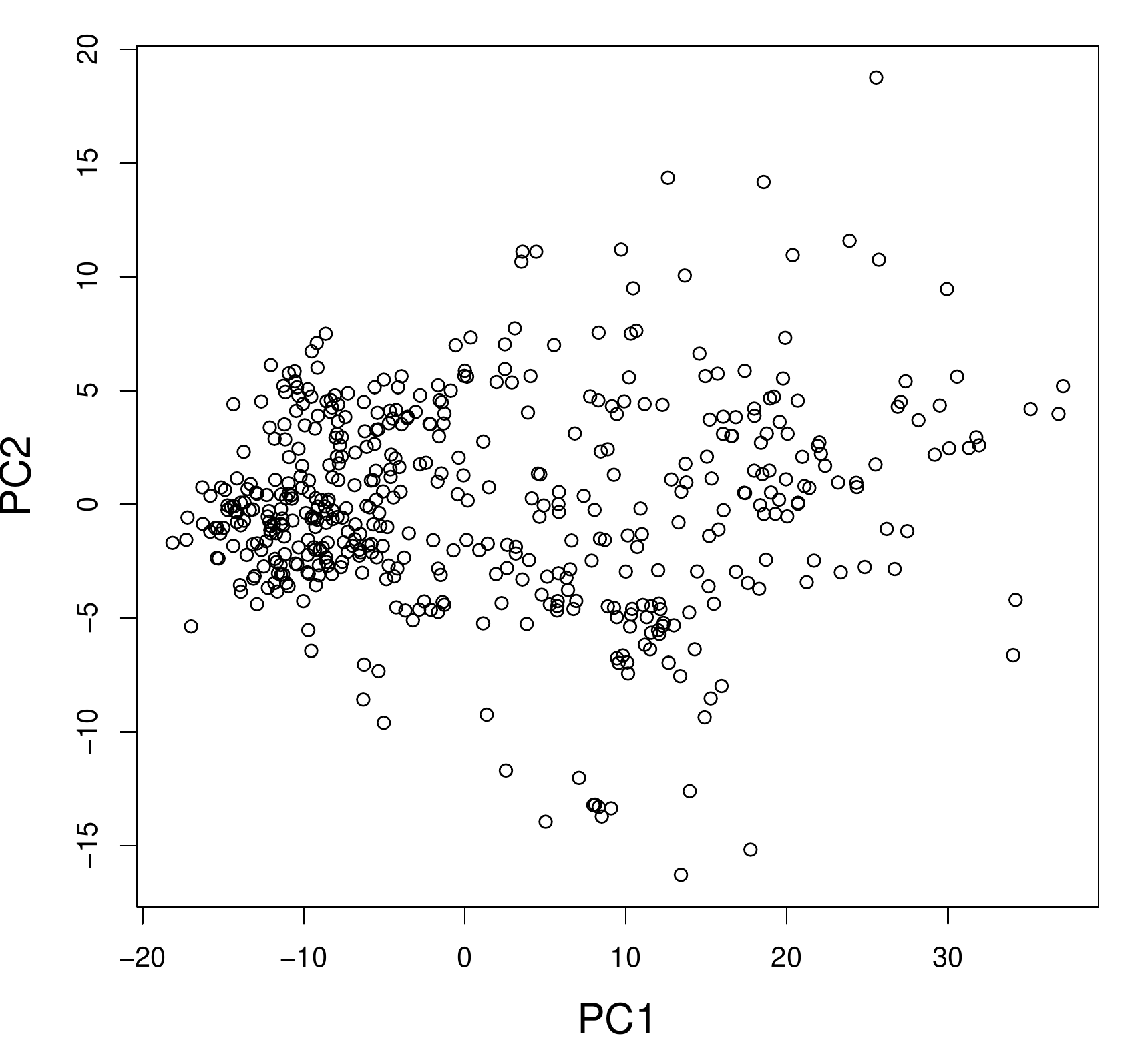}
\caption{The first two principal
   components of the cows data.} 
\label{fig:cows}
\end{figure}

The other real data example in the paper
is the cows dataset with 4 variables.
Our first instinct was to carry out a
PCA to get some idea about the shape of 
the data.
Figure \ref{fig:cows} shows the first
two principal components of the cows 
data, which explain 96\% of the total 
variance. (Here we used the ROBPCA
method of \cite{Hubert:ROBPCA},
but classical PCA gave a very similar 
picture.)
The plot shows that this dataset is
equally challenging. 
CRAC interpret it as a well-behaved
point cloud plus contamination.
Again most of the contamination is 
nearby, and then it fans out.
An alternative interpretation is that 
it could be a sample from a skewed 
distribution, and in that model none 
of these points need to be considered 
outlying.

The paper starts by analyzing the geyser
data through monitoring an S-estimator, 
the subsequent MM-estimator, and the 
MCD. Each of these estimators is tuned 
by its breakdown value.
For the S-estimator this yields the 
surprising conclusion that tuning for a 
50\% breakdown value achieves the 
intended result, whereas 49\% does not.
A second conclusion is that the tuning of 
the MM-estimator appears more stable, but 
that is only because the MM-estimator 
starts from the S-estimator with 50\% 
breakdown value, and it fails when 
starting from the 49\% version.

These conclusions would appear to 
suggest that S-estimators are not 
suitable in such challenging situations. 
In fact, using the Matlab code kindly
provided by CRAC we carried out an 
additional experiment, confirming that
if the small cluster is moved a little
bit in the direction of the larger one, 
even the S-estimator tuned for 50\% 
breakdown no longer detects it,
resulting in an uninformative 
monitoring plot consisting of 
horizontal lines only.
Following the S-estimate by the MM 
step does not change that of course.

In both the geyser and cows examples,
as well as the two simulated data sets
in Section 6 whose contamination is
equally nearby, CRAC find that the
MCD gives better results due to its
hard rejection.

\section{The choice of 
  the rho function}

We would like to argue that it is not
the definition of S-estimators that
makes them fail the stated
objective in these four examples, but 
rather the choice of the
$\rho$-function.
In our notation, a multivariate 
S-estimator 
of a $p$-variate dataset
$\{\by_1,\ldots,\by_n\}$
is defined as the pair $(\bT,\bC)$
formed by a location vector $\bT$ and
a PSD scatter matrix $\bC$ that 
together minimize $\mbox{Det}(\bC)$ 
subject to
\begin{equation}
\frac{1}{n}\sum_{i=1}^{n}
  {\rho(d_i(\bT,\bC))} = K
\end{equation}
in which 
$d_i(\bT,\bC) = \sqrt{
 (\by_i-\bT)'\bC^{-1}(\by_i-\bT)}$
is the statistical distance of 
$\by_i$ to $\bT$ relative to $\bC$.
One often takes
  $K = E[\rho(||\bZ||)]$ 
where $\bZ$ follows the multivariate 
standard normal distribution.
Note that we write $\rho$ as a function 
of the statistical distance $d_i$ and
not its square $d_i^2$\,.

People almost exclusively use the 
bisquare $\rho$ in S-estimators, which 
redescends slowly in order to attain 
a high statistical efficiency at 
uncontaminated data.
This is what makes it so hard to detect 
nearby contamination.
To illustrate our point, we construct
a `custom made' $\rho$-function
for dealing with nearby outliers.
It is given by
\begin{equation}\label{eq:rhof}
\rho_{a}(d) = 
\begin{cases}
  d^2/2 & 
	\mbox{ if } 0 \ls d \ls \sqrt{p} \\
  \left((1+a)\left(2\sqrt{p}d -
	  p\right)-d^2\right)/(2a) &  
	\mbox{ if } \sqrt{p} < d \ls 
	 (1+a)\sqrt{p}\\
  p(1+a)/2 & 
  \mbox{ if } d > (1+a)\sqrt{p} 
\end{cases}
\end{equation}
where $a \gs 0$ is a tuning 
constant that determines where the 
$\rho$-function becomes flat. Note 
that the dimension $p$ matters, because 
for uncontaminated Gaussian data the 
squared distance roughly follows a
$\chi^2$ distribution with $p$
degrees of freedom which has
expectation $p$.
The corresponding $\psi$-function is 
given by 
$\psi_a(d) = \rho_a'(d)$\,: 
\begin{equation}\label{eq:psif}
\psi_{a}(d) = 
\begin{cases}
d & \mbox{ if } 0 \ls d \ls \sqrt{p}\\
 \left(\sqrt{p}(1+a) - d \right)/a &
 \mbox{ if } 
   \sqrt{p} < d \ls (1+a)\sqrt{p}\\
  0 & \mbox{ if } d > (1+a)\sqrt{p}  
\end{cases}
\end{equation}
and the weight function by  
 $w_a(d) = \psi_a(d)/d$\,:
\begin{equation}\label{eq:wf}
w_{a}(d) = 
\begin{cases}
1 & \mbox{ if } 0 \ls d \ls \sqrt{p}\\
 \left(\sqrt{p}(1+a) - d \right)/(ad) & 
 \mbox{ if }
   \sqrt{p} < d \ls (1+a)\sqrt{p}\\
  0 & \mbox{ if } d > (1+a)\sqrt{p}
	\;\;. 
\end{cases}
\end{equation}
For this $\rho$-function, 
$K = E[\rho(||Z||_2)]$ can be computed 
numerically or by Monte Carlo.

\begin{figure}[!ht]
\centering
\includegraphics[width = 1.0\linewidth]
  {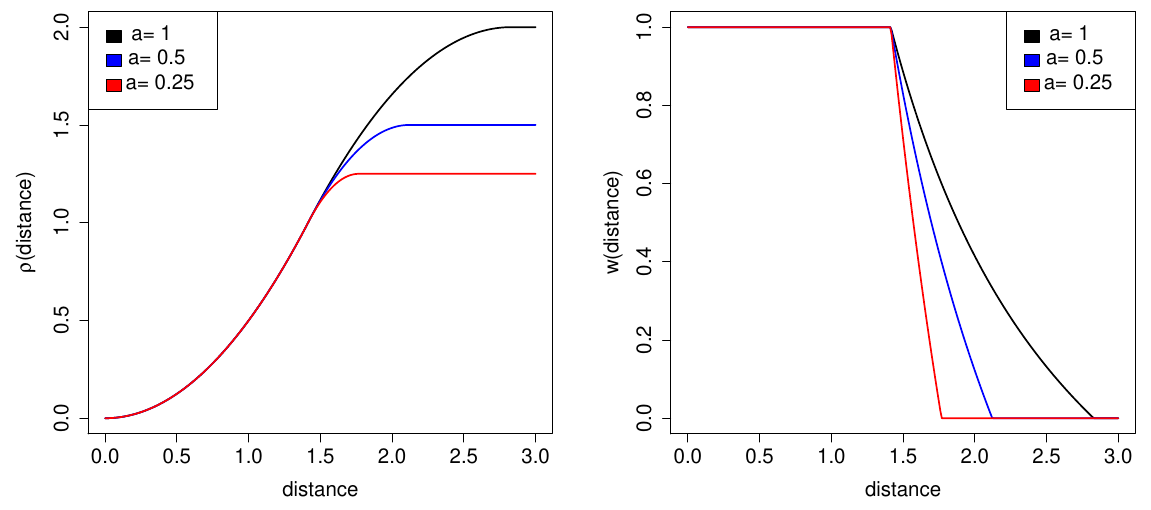}	
\caption{The custom made $\rho$-function
  $\rho_{a}$ (left) and the corresponding
	weight function $w_{a}$ (right) for 
  $p=2$, for different values of $a$.}
\label{fig:rhoweight}
\end{figure}
	
Figure \ref{fig:rhoweight} shows the
functions $\rho_{a}$ and $w_{a}$ for 
dimension $p=2$, for different values 
of $a$. In the right panel we see that 
the more $a$ approaches 0, the harder 
the rejection becomes.

\begin{figure}[!ht]
\centering
\includegraphics[width = 1.0\linewidth]
  {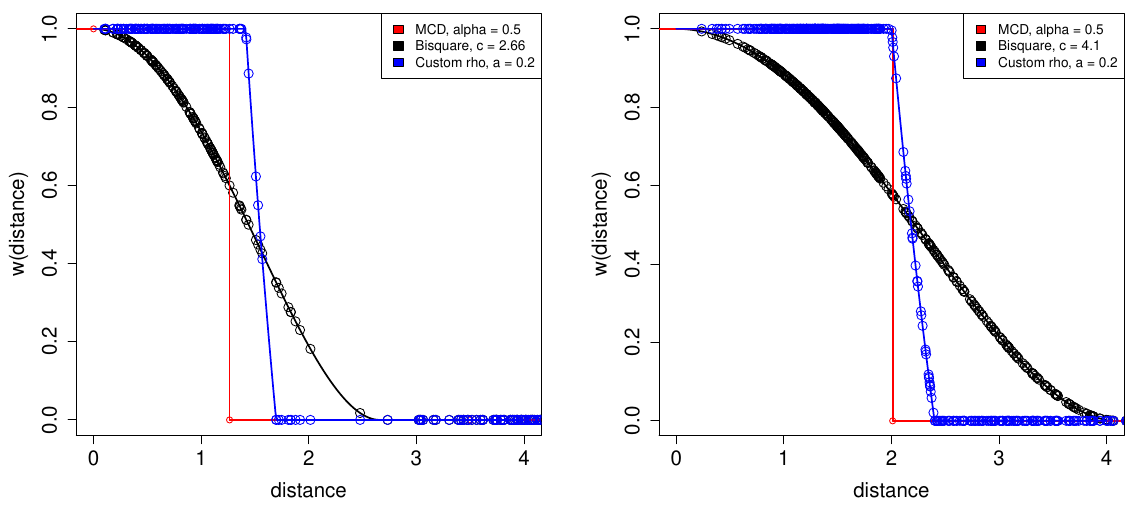}	
\caption{Comparing the distance weights
  of the MCD estimator and the scatter 
	S-estimators with bisquare 
	$\rho$-function and custom $\rho_{a}$
	for the geyser data (left) and the
	cows data (right).}	
\label{fig:weightcomps}
\end{figure}

The left panel of Figure 
\ref{fig:weightcomps} shows the weights 
assigned to the observations in the geyser 
data, where the statistical distances on
the horizontal axis were computed from
the MCD estimates of location and scatter.
The MCD method (red line) assigns weights
that are one for most points of the
main cluster and zero for all points of
the second cluster.
In contrast, the bisquare (black curve)
still gives quite a bit of weight to the 
points lying in between the two clusters.
(The weights are shown as little circles
on the curve.)
These points pull the S-estimator towards
a non-robust solution. 
Our custom weight function (blue curve)
redescends much faster so it gives small
weights to the intermediate points,
thereby behaving more like the MCD. 
The right panel shows the weights
in the cows data, with similar
conclusions.

\begin{figure}[!ht]
\centering
\includegraphics[width = 0.7\linewidth]
   {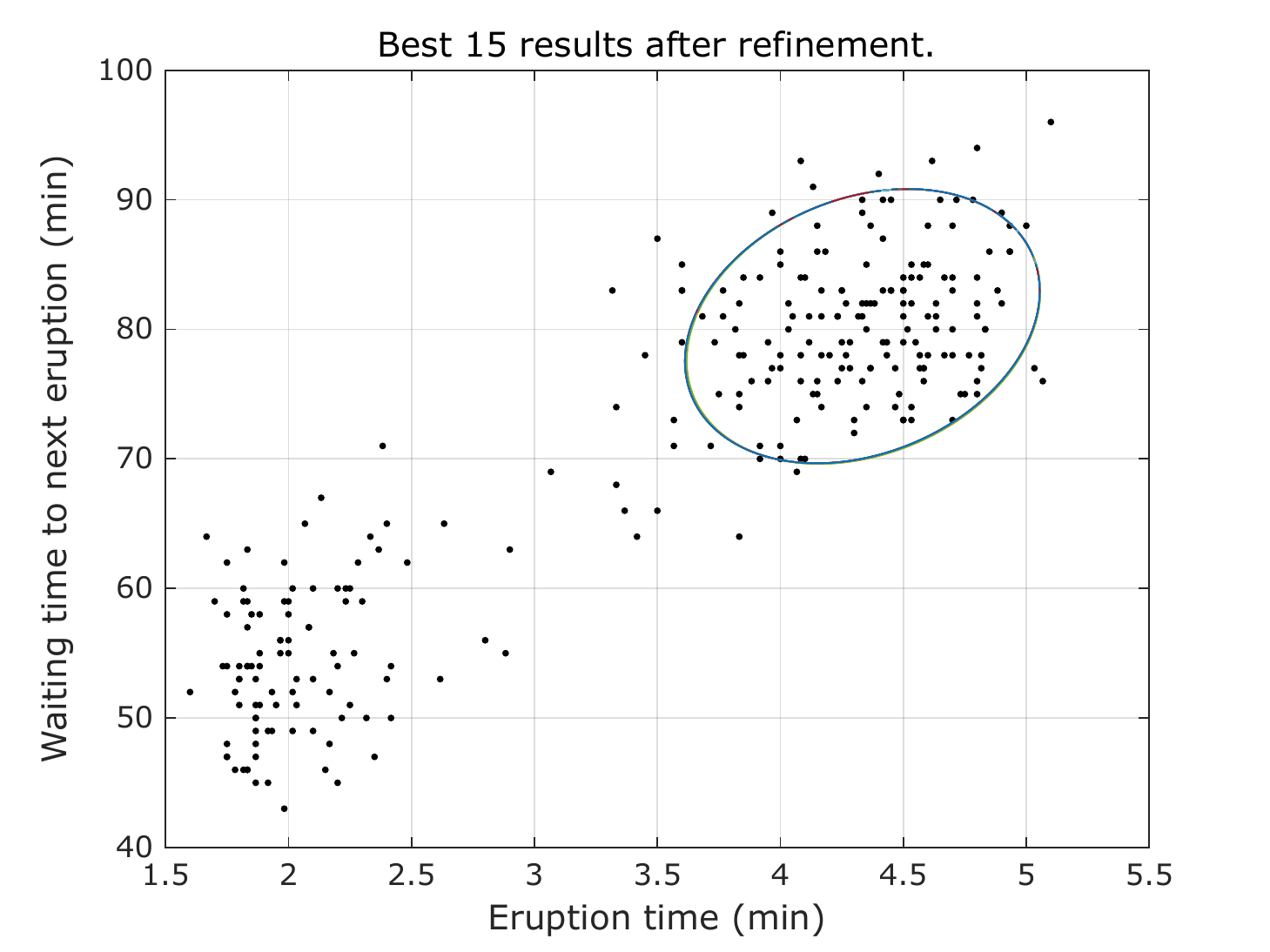}
\caption{Geyser data: fit obtained by
  the S-estimator with the custom 
	$\rho_a$ function.}
\label{fig:custS}
\end{figure}

We applied the S-estimator with the
custom function $\rho_a$ for $a = 0.2$ 
to the geyser dataset, using the FSDA
toolbox of \cite{Riani:FSDA}.
This does yield a fit to the main cluster, 
as illustrated by the tolerance ellipse in 
Figure \ref{fig:custS}.
For the cows data we also obtain the 
desired fit. 

We constructed this particular custom 
function $\rho_a$ with the sole purpose 
of illustrating that the behavior of 
S-estimators does not only depend 
on tuning, but also on the shape of 
the $\rho$-function.
The shape of this particular $\rho_a$
makes the S-estimator use weights
similar to those of the MCD, 
especially when $a$ is tiny. 
This exemplifies the tradeoff between
the ability to deal with nearby outliers
and statistical efficiency at Gaussian
data.
But we definitely do not propose 
$\rho_a$ for general use in multivariate
S-estimators because it has some
disadvantages. If we use 
$K = E[\rho(||\bZ||)]$ its breakdown
value $\min(K/\max(\rho_a),
1 - K/\max(\rho_a))$ is below the
50\% of the MCD and decreases with 
the dimension $p$. We could also
put $K=\max(\rho_a)/2$ but then we
obtain an inconsistent estimator that 
needs a consistency correction. 
In either case we do not know how many 
data points will fall in the 
non-constant part of $\rho_a$. 
For the MCD this number is always $h$, 
and the MCD is also easier to compute.

\section{Speeding up the computation}
Monitoring requires the repeated 
calculation of a statistical procedure 
for various values of the parameter that 
needs tuning. 
Therefore the feasibility of monitoring
depends on the computation time of the
method being monitored. 
The Matlab code used in the paper
computes the S-estimator, MM-estimator 
and MCD from 1000 initial elemental 
subsets that are kept unchanged
throughout the monitoring.
Computing concentration steps from 1000 
subsets is computationally demanding.
An alternative is to run the 
deterministic algorithm of 
\cite{Hubert:DetMCD} who start from 
six specifically constructed initial 
estimates instead of 1000 random ones,
and illustrated their method by 
monitoring memberships in a multivariate
classification.
\cite{Hubert:DetS} extended this work
to S-estimators and MM-estimators and
illustrated it by monitoring the
estimates of location and scatter in
a flow cytometry dataset with 
$n=29,486$ data points.

\section{Limitations of monitoring}
The review paper by CRAC clearly 
documents the benefits of monitoring.
On the other hand the approach also has
some limitations having to do with the
size of the problem and the
number of parameters that should be 
tuned.
One of us works in the food sorting
industry where thousands of items are 
inspected in milliseconds in order to 
detect outliers (impurities), yielding
settings with tens of dimensions.
In such situations monitoring could 
still be of use for tuning 
classification parameters, as 
long as it can be done off-line. 
The large sample size is the least
problematic: instead of monitoring all 
the individual observations one can 
monitor a subset of them, or summary
statistics.
Higher dimensions are more difficult
to handle due to the substantially 
increasing computation time.
Another question is how many parameters
can be tuned, as the parameters 
usually interact with each other. The 
paper illustrates tuning one parameter, 
and if there are more the experiment 
needs to be carefully designed.

\end{document}